\documentclass[12pt,draftcls, onecolumn]{IEEEtran}

\usepackage{amsmath}
\usepackage{amssymb}
\usepackage[dvips]{graphicx}
\usepackage{setspace}
\usepackage{epsfig}
\usepackage{amsmath}
\usepackage{amssymb}
\usepackage[dvips]{graphicx}
\usepackage{epsfig}

\newcommand{\p}{\overline{P}}
\newcommand{\tSNR}{\text{SNR}}
\newcommand{\E}{\mathbb{E}}
\newtheorem{theo}{Theorem}
\newcommand{\figsize}{0.65}

\begin{document}
\title{Performance Analysis of Cognitive Radio Systems under QoS Constraints and Channel Uncertainty}
\author{
\vspace{.2cm}
\authorblockN{Sami Akin and
Mustafa Cenk Gursoy}
\thanks{The authors are with the Department of Electrical
Engineering, University of Nebraska-Lincoln, Lincoln, NE, 68588
(e-mails: samiakin@huskers.unl.edu, gursoy@engr.unl.edu).}
\thanks{This work was supported by the National Science Foundation under
Grants CCF -- 0546384 (CAREER), CNS -- 0834753, and CCF--0917265.}}
\date{}

\maketitle

\begin{abstract}
In this paper, performance of cognitive transmission over time-selective flat fading channels is
studied under quality of service (QoS) constraints and channel
uncertainty. Cognitive secondary users (SUs) are assumed to initially perform channel sensing to detect the activities of the primary users, and then attempt to estimate the channel fading coefficients through training. Energy detection is employed for channel sensing, and different minimum mean-square-error (MMSE) estimation methods are considered for channel estimation. In both channel sensing and estimation, erroneous decisions can be made, and hence, channel uncertainty is not completely eliminated. In this setting, performance is studied and interactions between channel sensing and estimation are investigated.

Following the channel sensing and estimation tasks, SUs engage in data transmission. Transmitter, being unaware of the channel fading coefficients, is assumed to send the data at fixed power and rate levels that depend on the channel sensing results. Under these assumptions, a state-transition model is constructed by considering the reliability of the transmissions, channel sensing decisions and their correctness, and the evolution of primary user activity which is modeled as a two-state Markov process. In the data transmission phase, an average power constraint on the secondary users is considered to limit the interference to the primary users, and statistical limitations on the buffer lengths are imposed to take into account the QoS constraints of the secondary traffic.  The maximum throughput under these statistical QoS constraints is identified by finding the effective capacity of the cognitive radio channel. Numerical results are provided for the power and rate policies.


\end{abstract}

\newpage

\setcounter{page}{1}

\begin{spacing}{1.48}
\section{Introduction}
Cognitive radios have recently been studied intensively as they provide strategies to use the transmission spectrum more efficiently by enabling the cognitive secondary users (SUs) to use the transmission bands allocated to the licensed primary users (PUs) while causing no or only limited (or tolerable) interference to them. Interference is controlled by having the cognitive SUs be aware of the environment (e.g., through channel sensing) and adapt their transmission strategies accordingly. An overview of cognitive radio systems and the challenges in this area can be found in \cite{Mitola}-\cite{Quan}.

As discussed above, the central challenge for the cognitive SUs is to control their interference levels. In general, interference management needs to be performed under uncertainty as channel sensing done by the SUs may result in false alarms and miss-detections. In such an interference limited scenario, cognitive SUs should also satisfy their own quality of service (QoS) requirements by transmitting at high rates and limiting the delay experienced by the data in the buffers. This, too, has to be achieved under channel uncertainty since wireless channel conditions, which vary over time randomly due to mobility and changing environment, can only be estimated imperfectly through training techniques. Note also that providing QoS guarantees is especially more challenging for SUs as they have to take into account both the changing channel conditions and varying primary user activity. These considerations are critical for the successful deployment of cognitive radio systems in practice. Motivated by this, we in this paper study the performance of cognitive transmissions in a practical scenario in which cognitive SUs perform channel sensing and channel estimation, and operate under QoS constraints and also uncertainty caused by erroneous decisions in sensing and estimation steps. Below, we delineate the operation of the cognitive SUs.

Initially, before using the channel, SUs have to detect the activities of the primary users. Among different channel detection techniques, sensing-based access to the channel is favored because of its low employment cost and compatibility with the legacy of licensed systems \cite{Ghasemi}. The authors in \cite{Zhao2} and \cite{Zhao3} developed an optimal strategy for opportunistic spectrum access. Moreover, the authors in \cite{Jiang} focused on the optimal sensing order problem in multi-channel cognitive medium access control with opportunistic transmission, and studied the problem of maximally utilizing the spectrum opportunities in cognitive radio networks with multiple potential channels.


In wireless communications, as discussed above, channel conditions vary over time, and estimation of these channel variations is a crucial task before data transmission is performed. If the channel conditions are not known a priori, generally practical wireless systems employ training sequences to perform channel estimation. One of the early studies on channel training was conducted by Cavers who provided an analytical approach to the design of pilot-assisted modulations in \cite{Cavers1} and \cite{Cavers2}. These pilot-assisted transmission (PAT) strategies, which multiplex known training symbols with the data symbols, can be used for channel estimation, receiver adaptation, and optimal decoding \cite{Tong}. We assume that, following channel sensing, SUs perform channel estimation to learn the channel conditions. Due to interactions and interdependencies between channel sensing and estimation, we are faced with a challenging scenario. For instance, not detecting the activities of primary users reliably can lead to degradations in the estimation of the channel conditions, e.g., if the primary users are active but detected as idle, the quality of the channel estimate will deteriorate.


After performing the sensing and estimation tasks, SUs initiate the data transmission phase. We assume that SUs operate under QoS constraints in the form of limitations on the buffer length. In order to identify the maximum throughput under such constraints, we employ the effective capacity as a performance metric \cite{Wu}. Recently, effective capacity analysis of wireless systems has attracted much interest (see e.g., \cite{Liu} and \cite{zhang3}). In \cite{Ak_Gur1}, we studied the cognitive transmission under quality of service (QoS) constraints. In \cite{Ak_Gur2}, by initially performing channel sensing over multiple frequency bands to detect the activities of primary users, we studied the performance limits of cognitive transmission under QoS constraints and interference limitations. In this work, SUs are assumed to have perfect knowledge of the channel conditions.

In this paper, considering that no prior channel knowledge is available at the secondary transmitter and the secondary receiver, we study the effective capacity of cognitive radio channels in order to identify the performance limits under channel uncertainty and QoS constraints. The cognitive radio is assumed to initially perform channel sensing, and then estimate the channel fading coefficients, and finally perform data transmission. The activity of primary users is modeled as a two-state  Markov process\footnote{In addition to having the assumption of no prior channel knowledge and explicitly considering channel estimation, Markovian modeling of primary user activity constitutes another significant departure from the setting considered in \cite{Ak_Gur2} where primary user activity is assumed to vary independently from one frame to another.}. In this setting, we jointly optimize the training symbol power, data symbol power and transmission rates. More specifically, the contributions of this paper are the following:
\begin{enumerate}
  \item We jointly consider channel sensing and estimation, and provide a framework through which the performance of cognitive transmissions can be analyzed in the presence of uncertainty caused by errors in sensing and estimation.

  \item We identify a state-transition model for cognitive transmission by modeling the primary user activity as a two-state Markov process, considering the reliability of the transmissions, and taking into account the sensing and channel estimation decisions and their correctness.

  \item We determine the effective capacity of cognitive transmissions and obtain the maximum throughput under QoS constraints and channel uncertainty.

  \item We numerically analyze the power and rate allocation strategies and identify the maximum throughput levels as a function of detection and false alarm probabilities and training power.

\end{enumerate}

The organization of the rest of the paper is as follows. In Section \ref{CChannel Model}, we describe the cognitive channel model. In Section \ref{Channel Sensing}, we discuss channel sensing and provide expressions for the probability of detection and false alarm. In Section \ref{PSAM State Tra}, we describe channel training with pilot symbols and discuss different channel estimation techniques. In Section \ref{sec:state}, we construct a state transition model for cognitive radio transmission. In Section \ref{Effective Capacity sec}, we identify the maximum throughput that the cognitive radio channel can sustain under QoS constraints. In Section \ref{Numerical Results}, we provide the numerical results. We conclude in Section \ref{Conclusion}.

\section{Cognitive Channel Model}\label{CChannel Model}
\begin{figure}
\begin{center}
\includegraphics[width =\figsize\textwidth]{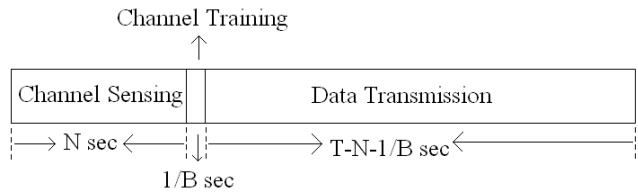}
\caption{Transmission frame consisting of channel sensing, channel training and data transmission. Total frame duration is $T$. First $N$ seconds is allocated to channel sensing. Following channel sensing, a single pilot symbol is sent in the training phase. Under the assumption that the symbol rate is $B$ complex symbols per second, a single pilot has a duration of $1/B$ seconds, where $B$ denotes the bandwidth. The remaining time of $T -N - 1/B$ seconds is used for data transmission.} \label{fig:Resim_extra}
\end{center}
\end{figure}

We consider a cognitive channel model in which a secondary
transmitter sends information to a secondary receiver. Initially, the secondary users perform channel sensing. Then, depending on the channel sensing results, the secondary transmitter selects pilot symbol and data transmission power policy. Note that the pilot symbol is used for the estimation of the channel fading coefficients. We assume that channel sensing, channel estimation, and data transmission are performed in frames of total duration $T$ seconds. In each frame, the first $N$ seconds is allocated for channel sensing. Following the channel sensing, a single pilot symbol is employed to enable the secondary receiver to estimate the channel fading coefficient. Then, data transmission is performed. The allocation of the frame duration to these tasks is depicted in Figure \ref{fig:Resim_extra}.

Pilot symbol and data symbol powers, and transmission rates depend on the channel sensing results, i.e., if the channel is sensed to be busy (correct detection of busy case or false alarm), the average transmission power and transmission rate are set to $\overline{P}_{1}$ and $r_{1}$, respectively. If, on the other hand, the channel is sensed to be idle (misdetection or correct detection of idle case), the average transmission power and transmission rate are set to $\overline{P}_{2}$ and $r_{2}$, respectively. Note that if $\overline{P}_{1}=0$, the secondary transmitter stops the transmission when the primary users are sensed to be active.

The discrete-time channel input-output relation between the
secondary transmitter and receiver in the $i^{th}$ symbol duration is given by
\begin{equation}\label{input-output 1}
y_{i}=h_{i}x_{i}+n_{i}\quad i=1,2,...,
\end{equation}
if the primary users are inactive. On the other hand, if the primary users are using the channel, we have
\begin{equation}\label{input-output 2}
y_{i}=h_{i}x_{i}+n_{i}+s_{i}\quad i=1,2,...,
\end{equation}
where $x_{i}$ and $y_{i}$ denote the complex-valued channel input and output, respectively. In (\ref{input-output 1}) and
(\ref{input-output 2}), $h_{i}$ represents the fading coefficient between the secondary transmitter and receiver. The fading coefficients are zero-mean, circularly symmetric, complex Gaussian distributed with variance $\mathbb{E}\{|h_{i}|^{2}\}=\sigma_{h}^{2}$. In (\ref{input-output 1}) and (\ref{input-output 2}), $\{n_{i}\}$ is a sequence of additive thermal random noise samples at the secondary receiver, that are zero-mean, circularly symmetric, complex Gaussian distributed with variance $\mathbb{E}\{|n_{i}|^{2}\}=\sigma_{n}^{2}$ for all $i$.
In (\ref{input-output 2}), $s_{i}$ denotes the sum of active primary users' faded signals arriving at the secondary receiver. We denote the variance of $s_{i}$ as $\sigma_s^2$. Note also that since the bandwidth is $B$, symbol rate is assumed to be $B$ complex symbols per second.

We consider block-fading and assume that the fading coefficients $\{h_i\}$ stay constant within each frame of $T$ seconds and change independently from one frame to another. We also assume that the activity of the primary users stay the same in each frame. However, a two-state Markov model is employed to model the transitions of the activity of the primary users between the frames.

\section{Channel Sensing}\label{Channel Sensing}
Energy-detection methods are considered to be well-suited for channel sensing if the transmission policies of primary users are not known. We can formulate the channel sensing as a hypothesis testing problem between the noise $n_{i}$ and the signal $s_{i}$ in noise. Since the bandwidth is $B$, there are $NB$ complex symbols in a duration of $N$ seconds. Now, the hypothesis testing problem can mathematically be expressed as follows:
\begin{align}\label{hypothesis}
\begin{split}
&\mathcal{H}_{0}:y_{i}=n_{i}, \quad i=1,2,...,NB\\
&\mathcal{H}_{1}:y_{i}=n_{i}+s_{i}, \quad i=1,2,...,NB.
\end{split}
\end{align}
We assume that $s_{i}$ has a circularly symmetric complex Gaussian distribution\footnote{Note that if the signals are being received in a rich multipath environment or the number of active primary users is large, the simplifying Gaussian assumption for the distribution of $s_{i}$ has high accuracy. Moreover, if, for example the primary users are employing frequency or phase modulation, $s_{i}$ in the presence of even a single primary user in flat fading Rayleigh channel will be Gaussian distributed.} with zero-mean and variance $\sigma_{s}^{2}$. Furthermore, as in \cite{liang}, we assume that the signal samples $\{s_{i}\}$ are independent and identically distributed (i.i.d.). Under these assumptions, the optimal Neyman-Pearson detector for the above hypothesis problem is given by \cite{Poor-book}
\begin{equation}\label{Neyman-Pearson}
Y=\frac{1}{NB}\sum_{i=1}^{NB}|y_{i}|^{2}\gtrless_{\mathcal{H}_{0}}^{\mathcal{H}_{1}}\lambda
\end{equation}
where $\lambda$ is the detection threshold. Observing that $Y$ is chi-squared distributed with $2NB$ degrees of freedom, we can establish the probabilities of false alarm and detection as follows:
\begin{align}\label{false alarm}
&P_{f}=\Pr\{Y>\lambda\mid\mathcal{H}_{0}\}=1-P\left(\frac{NB\lambda}{\sigma_{n}^{2}},NB\right)\\
&P_{d}=\Pr\{Y>\lambda\mid\mathcal{H}_{1}\}=1-P\left(\frac{NB\lambda}{\sigma_{n}^{2}+\sigma_{s}^{2}},NB\right)\label{eq:probdetect}
\end{align}
where $P(x,y)$ denotes the regularized gamma function and is defined as $P(x,y)=\frac{\gamma(x,y)}{\Gamma(y)}$ where $\gamma(x,y)$ is the lower incomplete gamma function and $\Gamma(a)$ is the Gamma function.

Above, we have considered an i.i.d. scenario. If $\{s_{i}\}$ are correlated and if the correlation structure is known by the cognitive users, then the optimal detector computes, as the test statistic, the quadratic form $\mathbf{y}^\dagger\mathbf{K}\mathbf{y}$ where $\mathbf{y}$ is the vector of $NB$ received signal samples $\{y_{i}\}_{i = 1}^{NB}$, and $\mathbf{K}$ is a matrix that depends on the covariance matrix of the primary user signal samples $\{s_{i}\}_{i=1}^N$ \cite[Case III.B.4]{Poor-book}. If
$\{s_{i}\}$ are identically distributed, then the false alarm and detection probabilities are again expressed in terms of the regularized lower gamma function and are in the same form as in (\ref{false alarm}) and (\ref{eq:probdetect}) (see \cite[Equation III.B.96]{Poor-book}).

In the hypothesis testing problem given in (\ref{hypothesis}),
another approach is to consider $Y$ as Gaussian distributed, which is accurate if $NB$ is large \cite{liang}. In this case, the detection and false alarm probabilities can be expressed in terms of Gaussian $Q$-functions. We would like to note that the rest of the analysis in the paper does not depend on the specific expressions of the false alarm and detection probabilities. However, numerical results are obtained using (\ref{false alarm}) and (\ref{eq:probdetect}).

A similar hypothesis-testing formulation for channel sensing is also studied in \cite{Ak_Gur1} and is provided in this paper as well for the completeness of the discussion.

\section{Pilot Symbol-Assisted Transmission}\label{PSAM State Tra}
After channel sensing is performed, the secondary transmitter sends the pilot symbol to enable the receiver to estimate the channel fading coefficient. In this section, we consider several channel estimation methods. As emphasized earlier, channel estimation has dependence on channel sensing results. Regarding the channel sensing result and its correctness, we have the following four possible scenarios:
\begin{enumerate}
  \item \emph{Scenario 1:} Channel is busy, detected as busy (correct detection),
  \item \emph{Scenario 2:} Channel is busy, detected as idle (miss-detection),
  \item \emph{Scenario 3:} Channel is idle, detected as busy (false alarm),
  \item \emph{Scenario 4:} Channel is idle, detected as idle (correct detection).
\end{enumerate}
Note that the secondary transmitter sends data with average power $\overline{P}_{1}$ if the channel is sensed as busy, whereas the transmitter sends data with average power $\overline{P}_{2}$ if the channel is detected to be idle. Since fading stays constant in each frame, it is enough to send only one pilot symbol in each frame\footnote{Since MMSE estimation depends only on the pilot power and not on the number of pilot symbols, a single pilot symbol with optimized power is sufficient.}. Therefore, the first $N$ seconds of a frame duration $T$ seconds is spared to sense the channel, a single pilot symbol is sent following channel sensing, and $(T-N)B-1$ data symbols are transmitted after the pilot symbol\footnote{Since the symbol rate is $B$ symbols per second, we have $(T-N)B$ symbols in a duration of $T-N$ seconds. Among these symbols, the first symbol is a pilot symbol and the remaining $(T-N)B - 1$ symbols are the data symbols.}. In each frame, the average input power is
\begin{equation}\label{power constraint}
\frac{1}{T}\sum_{i=(lT+N)B}^{(l+1)TB-1}\E\left\{|x_{i}|^{2}\right\} =
\overline{P}_{1} \quad l=0,1,2,...,
\end{equation}
when activity is sensed in the
channel, whereas the average input power is
\begin{equation}\label{power constraint}
\frac{1}{T}\sum_{i=(lT+N)B}^{(l+1)TB-1}\E\left\{|x_{i}|^{2}\right\} =
\overline{P}_{2} \quad l=0,1,2,...,
\end{equation}
when the channel is sensed to be idle. Above, $l$ denotes the frame index. From the average power constraints, we see that the total energy allocated to the pilot and data symbols is limited in one frame by $T\overline{P}_{1}$ or $T\overline{P}_{2}$ when the channel is busy or idle, respectively.

We assume that, depending on the the capabilities of the transmitters and the energy resources they are equipped with, there exists peak constraints on the average powers, e.g.,
\begin{gather}
\p_1 \le \p_{peak} \quad \text{ and } \p_2 \le \p_{peak}.
\end{gather}

Additionally, note that the secondary transmitter transmits with an average power $\overline{P}_{1}$ in scenario 1 and with an average power $\overline{P}_{2}$ in scenario 2. In both scenarios,  primary users are active in the channel and experience interference due to the transmission of the secondary users.
In order to limit the interference and protect the primary users, we impose the following constraint on $\overline{P}_{1}$ and $\overline{P}_{2}$:
\begin{equation}\label{general power constraint}
P_{d}\overline{P}_{1}+(1-P_{d})\overline{P}_{2}\leq
\overline{P}_{avg}
\end{equation}
where $P_{d}$ is the probability of detection and $(1-P_{d})$ is the probability of miss-detection, and $P_d$ and $1-P_d$ can be regarded as the probabilities of scenarios 1 and 2, respectively. In the following, we describe how $\p_{avg}$ can be related to the interference constraints. Let us denote the fading coefficient between the secondary transmitter and primary receiver as $h_{sp}$. Now, the average interference experienced by the primary user can be expressed as
\vspace{-.3cm}
\begin{gather} \label{eq:averageinterference}
\E\{ P_{d}\overline{P}_{1} |h_{sp}|^2 + (1-P_{d})\overline{P}_{2} |h_{sp}|^2\} =  \left(P_{d}\overline{P}_{1} + (1-P_{d})\overline{P}_{2} \right) \E\{|h_{sp}|^2\} \le I_{avg}
\end{gather}
where $I_{avg}$ can be regarded as the average interference constraint. We assume that the realizations of $h_{sp}$ are not known at the secondary transmitter and hence the secondary transmitter cannot adapt its transmission according to $h_{sp}$. However, if the statistics of $h_{sp}$ (e.g., the mean of $|h_{sp}|^2$ is known), then the secondary transmitter can choose $P_{avg} = \frac{I_{avg}}{\E\{|h_{sp}|^2\}}$ in order to satisfy (\ref{eq:averageinterference}).


Finally, we would like to note that in the perfect detection case in which $P_d = 1$, there are no miss-detections and (\ref{general power constraint}) specializes to $\p_1 \le \p_{avg}$. Hence, expectedly, only $\p_1$, which is the transmission power when the channel is sensed as busy, is affected by the interference constraints, and we have $\p_2 \le \p_{peak}$. If $P_d < 1$, miss-detections should also be considered. In such cases, the secondary users do not detect the active primary users and transmit at power $\p_2$. Hence, $\p_2$ should also be considered in interference control as formulated in (\ref{general power constraint}).

\subsection{Training Phase} \label{subsec:training}
In the channel training phase, the pilot symbol power is set depending on the sensing result. If the channel is detected
as busy, the power of pilot symbol is set to $P_{tb}=\eta\overline{P}_{1}T$. On the other hand, the pilot power is $P_{ti}=\eta\overline{P}_{2}T$ when no activity is detected. $\eta$ is the fraction of the total power allocated to the pilot symbol.

For the scenarios described at the beginning of this section, the corresponding received signals in the training phase are given by the following:
\begin{enumerate}
\item \emph{Scenario 1:} $y_{(lT+N)B}=h_{l}\sqrt{P_{tb}}+n_{(lT+N)B}+s_{(lT+N)B}$,
\item \emph{Scenario 2:} $y_{(lT+N)B}=h_{l}\sqrt{P_{ti}}+n_{(lT+N)B}+s_{(lT+N)B}$,
\item \emph{Scenario 3:} $y_{(lT+N)B}=h_{l}\sqrt{P_{tb}}+n_{(lT+N)B}$,
\item \emph{Scenario 4:} $y_{(lT+N)B}=h_{l}\sqrt{P_{ti}}+n_{(lT+N)B}$.
\end{enumerate}
Above, $h_{l}$ denotes the
channel fading coefficients in the $l^{th}$ block. The fading
coefficients are estimated via MMSE estimation, which provides the following estimates for each scenario:
\begin{align}
&\textit{Scenario 1:} \quad
\widehat{h}_{l, m-mmse}=\frac{\sqrt{P_{tb}}\sigma_{h}^{2}}{P_{tb}\sigma_{h}^{2}+\sigma_{n}^{2}+\sigma_{s}^{2}}y_{(lT+N)B}, \label{eq:estimate1}\\
&\textit{Scenario 2:} \quad
\widehat{h}_{l, m-mmse}=\frac{\sqrt{P_{ti}}\sigma_{h}^{2}}{P_{ti}\sigma_{h}^{2}+\sigma_{n}^{2}}y_{(lT+N)B},
\label{eq:estimate2}\\
&\textit{Scenario 3:} \quad
\widehat{h}_{l, m-mmse}=\frac{\sqrt{P_{tb}}\sigma_{h}^{2}}{P_{tb}\sigma_{h}^{2}+\sigma_{n}^{2}+\sigma_{s}^{2}}y_{(lT+N)B},
\label{eq:estimate3}\\
&\textit{Scenario 4:} \quad
\widehat{h}_{l,m-mmse}=\frac{\sqrt{P_{ti}}\sigma_{h}^{2}}{P_{ti}\sigma_{h}^{2}+\sigma_{n}^{2}}y_{(lT+N)B} \label{eq:estimate4}.
\end{align}
From above, we see that the estimate expressions in scenarios 1 and 3 in which the channel is detected as busy are the same. So are the expressions in scenarios 2 and 4 in which the channel is detected as idle. Hence, the receiver has two estimation rules depending on whether the channel is sensed as busy or idle.   Note that the MMSE formulation is obtained under the assumption that the primary users' signal $s$ is Gaussian distributed with mean zero and variance $\sigma_s^2$.
It is also important to note that the above MMSE estimates are affected by the channel sensing results. For instance, in scenario 2, the channel is busy but the receiver senses the channel as idle. Based on this sensing result, the receiver assumes that the noise variance is $\sigma_n^2$ rather than the actual value $\sigma_n^2 + \sigma_s^2$, and multiplies the observation $y_{(lT+N)B}$ by $\frac{\sqrt{P_{ti}}\sigma_{h}^{2}}{P_{ti}\sigma_{h}^{2}+\sigma_{n}^{2}}$ instead of $\frac{\sqrt{P_{ti}}\sigma_{h}^{2}}{P_{ti}\sigma_{h}^{2}+\sigma_{n}^{2} + \sigma_s^2}$. Hence, in the computation of the MMSE estimate, the receiver treats its channel sensing decision as the true decision. Hence, if the sensing decision is erroneous, the MMSE estimate is obtained for a mismatched channel. For this reason, we call these estimates as \emph{mismatched MMSE} estimates and use the subscript $m-mmse$. Note that from the receiver's perspective, the variance of the noise is random taking two possible values, $\sigma^2_n$ and $\sigma^2_n + \sigma_s^2$. In the presence of uncertainty in the noise statistics, the true MMSE estimate is given by the following result.

\begin{theo} \label{theo:MMSEestimate}
Given the channel sensing decision and the observation $y$ in the training phase, the receiver obtains the MMSE estimate through the following formulation:
\begin{align}
\widehat{h}_{mmse} = \E\{h | y\} = \Pr\{\sigma^2 = \sigma_n^2 \mid y\} \, \frac{\sqrt{P_{t}}\sigma_{h}^{2}}{P_{t}\sigma_{h}^{2}+\sigma_{n}^{2}}y_{} + \Pr\{\sigma^2 = \sigma_n^2 + \sigma_s^2 \mid y\} \, \frac{\sqrt{P_{t}}\sigma_{h}^{2}}{P_{t}\sigma_{h}^{2}+\sigma_{n}^{2}+\sigma_{s}^{2}}y_{}
\end{align}
where $\Pr\{\sigma^2 = \sigma_n^2 \mid y\} = \frac{\Pr\{\sigma^2 = \sigma_n^2\}f(y|\sigma^2 = \sigma_n^2)}{f(y)}$ and similarly $\Pr\{\sigma^2 = \sigma_n^2 + \sigma_s^2 \mid y\} = \frac{\Pr\{\sigma^2 = \sigma_n^2 + \sigma_s^2\}f(y|\sigma^2 = \sigma_n^2 + \sigma_s^2)}{f(y)}$. In the above formulation, we have
\begin{align}
\Pr\{\sigma^2 = \sigma_n^2\} &= \left\{
\begin{array}{ll}
\frac{a P_f }{aP_f + bP_d} & \text{if the channel is detected busy}
\\
\frac{a (1-P_f) }{a(1-P_f) + b(1-P_d)} & \text{if the channel is detected idle}
\end{array}\right. \label{eq:prior1}
\\
\Pr\{\sigma^2 = \sigma_n^2 + \sigma_s^2\} &= \left\{
\begin{array}{ll}
\frac{b P_d }{aP_f + bP_d} & \text{if the channel is detected busy}
\\
\frac{b (1-P_d) }{a(1-P_f) + b(1-P_d)} & \text{if the channel is detected idle}
\end{array}\right. \label{eq:prior2}
\\
f(y|\sigma^2 = \sigma_n^2) &= \frac{1}{\pi (P_t \sigma_h^2 + \sigma_n^2)} e^{-\frac{|y|^2}{P_t \sigma_h^2 + \sigma_n^2}} \label{eq:fy1}
\\
f(y|\sigma^2 = \sigma_n^2 + \sigma_s^2) &= \frac{1}{\pi (P_t \sigma_h^2 + \sigma_n^2 + \sigma_s^2)} e^{-\frac{|y|^2}{P_t \sigma_h^2 + \sigma_n^2 + \sigma_s^2}} \label{eq:fy2}
\\
f(y) &= \Pr\{\sigma^2 = \sigma_n^2\}f(y|\sigma^2 = \sigma_n^2) + \Pr\{\sigma^2 = \sigma_n^2 + \sigma_s^2\}f(y|\sigma^2 = \sigma_n^2 + \sigma_s^2) \label{eq:fyavg}
\end{align}
In (\ref{eq:prior1}) and (\ref{eq:prior2}), $P_d$ and $P_f$ denote the detection and false-alarm probabilities, respectively, and $a$ and $b$ are the transition probabilities in the two-state Markov model of the primary user activity (depicted in Fig. \ref{fig:fig1} and described in detail in Section \ref{sec:state}). Note also that $P_t$ denotes the power of the pilot symbol and is equal to $P_{tb}$ if the channel is detected busy and equal to $P_{ti}$ is the channel if detected idle.
\end{theo}

\emph{Proof:} See Appendix \ref{app:proofMMSEestimate}. \hfill $\square$

It can be immediately seen that as in the mismatched MMSE case, we again have two estimation rules depending on the channel sensing result. Note that the statistical characterization (e.g., finding the variance or more generally the distribution) of the MMSE estimate in Theorem \ref{theo:MMSEestimate} is a difficult task and can only be done through numerical analysis. It is also computationally intensive for the receiver to obtain this estimate. Another strategy is to obtain the linear MMSE estimate. Note that given the observation $y$, the linear MMSE estimate is given by
\begin{align}
\widehat{h}_{l-mmse} = \frac{\E\{hy^*\}}{\E\{|y|^2\}} y
&= \frac{\sqrt{P_t}\sigma_h^2}{\E\{|y|^2\}} y
\\
&= \frac{\sqrt{P_t}\sigma_h^2}{\Pr\{\sigma^2 = \sigma_n^2\}\E\{|y|^2 \mid \sigma^2 = \sigma_n^2 \} + \Pr\{\sigma^2 = \sigma_n^2 + \sigma_s^2\}\E\{|y|^2 \mid \sigma^2 = \sigma_n^2 + \sigma_s^2\}} y \nonumber
\\
&= \frac{\sqrt{P_t}\sigma_h^2}{\Pr\{\sigma^2 = \sigma_n^2\}(P_t \sigma_h^2 + \sigma_n^2) + \Pr\{\sigma^2 = \sigma_n^2 + \sigma_s^2\}(P_t \sigma_h^2 + \sigma_n^2 + \sigma_s^2)} y \label{eq:linearMMSE}
\end{align}
where $\Pr\{\sigma^2 = \sigma_n^2\}$ and $\Pr\{\sigma^2 = \sigma_n^2 + \sigma_s^2\}$ are given in (\ref{eq:prior1}) and (\ref{eq:prior2}). Similarly, as in Theorem \ref{theo:MMSEestimate}, $P_t$ is either equal to $P_{tb}$ or $P_{ti}$ depending whether the channel is detected as busy or idle.

It is interesting to note that if channel sensing results are perfect, i.e., $P_d = 1$ and $P_f = 0$, all estimation methods discussed above converge.

\subsection{Data Transmission Phase}

Now, we can express the fading coefficients as follows
\vspace{-.3cm}
\begin{equation}\label{coeff}
h_{l}=\widehat{h}_{l}+\widetilde{h}_{l}
\end{equation}
where $\widetilde{h}_{l}$ is the estimation error. Consequently, the input-output relationship in the data transmission phase of the $l^{th}$ frame can be written as
\vspace{-.3cm}
\begin{equation}\label{input-ouput-data1}
y_{i}=\widehat{h}_{l}x_{i}+\widetilde{h}_{l}x_{i}+n_{i}+s_{i}\qquad (lT+N)B+1\leq i \le (l+1)TB-1
\end{equation}
if the channel is busy, and
\vspace{-.3cm}
\begin{equation}\label{input-ouput-data2}
y_{i}=\widehat{h}_{l}x_{i}+\widetilde{h}_{l}x_{i}+n_{i}\qquad (lT+N)B+1\leq i \le (l+1)TB-1
\end{equation}
if the channel is idle.

Note that the mismatched MMSE estimates in (\ref{eq:estimate1}) -- (\ref{eq:estimate4}) and linear MMSE estimate in (\ref{eq:linearMMSE}) can be written as $\widehat{h} = K y$ where $K$ is a constant that depends on the channel sensing result and $y$ is the received signal in the training phase. Since $y$ is a Gaussian random variable, $\widehat{h}_{l}$ and
$\widetilde{h}_{l}$ are zero-mean circularly symmetric complex Gaussian random variables in these cases. The variance of the channel estimates is $\sigma_{\widehat{h}}^{2} = K^2 \E\{|y|^2\}$. In particular, we have the following variance expressions for the mismatched MMSE estimates in different scenarios:
\vspace{-.3cm}
\begin{align}
&\textit{Scenario 1:}\quad
\sigma_{\widehat{h}_{l, m-mmse}}^{2}=\frac{P_{tb}\sigma_{h}^{4}}{P_{tb}\sigma_{h}^{2}+\sigma_{n}^{2}+\sigma_{s}^{2}},\nonumber\\
&\textit{Scenario 2:}\quad
\sigma_{\widehat{h}_{l,m-mmse}}^{2}=\frac{P_{ti}\sigma_{h}^{4}}{\left(P_{ti}\sigma_{h}^{2}+\sigma_{n}^{2}\right)^{2}}\left(P_{ti}\sigma_{h}^{2}+\sigma_{n}^{2}+\sigma_{s}^{2}\right),\nonumber\\
&\textit{Scenario 3:}\quad
\sigma_{\widehat{h}_{l, m-mmse}}^{2}=\frac{P_{tb}\sigma_{h}^{4}}{\left(P_{tb}\sigma_{h}^{2}+\sigma_{n}^{2}+\sigma_{s}^{2}\right)^{2}}\left(P_{tb}\sigma_{h}^{2}+\sigma_{n}^{2}\right),\nonumber\\
&\textit{Scenario 4:}\quad
\sigma_{\widehat{h}_{l, m-mmse}}^{2}=\frac{P_{ti}\sigma_{h}^{4}}{P_{ti}\sigma_{h}^{2}+\sigma_{n}^{2}}.\nonumber
\end{align}
In all scenarios, the variance of the estimation error in both mismatch and linear MMSE can be written as
\vspace{-.3cm}
\begin{equation}\label{estimation error variance}
\sigma_{\widetilde{h}_{l}}^{2}=(1-2K\sqrt{P_{t}})\sigma_{h}^{2}+\sigma_{\widehat{h}_{l}}^{2}.
\end{equation}
Again, $P_{t}$ is either $P_{tb}$ or $P_{ti}$ depending on whether the channel is sensed as busy or idle, respectively. In true MMSE, since the estimate and error are uncorrelated, we have $\sigma_{\widetilde{h}_{l}}^{2} = \sigma_{h}^{2} -  \sigma_{\widehat{h}_{l}}^{2}$.

\section{State Transition Model}\label{sec:state}
 In this section, we construct a state-transition model for cognitive transmission. In order to identify this model, we first consider the transmission rates that can be supported by the channel. In the presence of channel uncertainty, it is generally difficult to characterize the channel capacity, which is the maximum transmission rate at which reliable communications can be established \cite{gursoy}. Therefore, most studies work with lower bounds on the channel capacity. One common technique employed in deriving an achievable rate expression is to regard the error in the channel estimate as another source of Gaussian noise. Since Gaussian noise is the worst uncorrelated noise, this assumption leads to a lower bound \cite{training}. On the other hand, the achievable rate expressions obtained using this approach are good measures of the rates supported in communication
systems that operate as if the channel estimate were perfect (i.e., in systems where Gaussian codebooks designed for known channels are used, and scaled nearest neighbor decoding is employed at the receiver) \cite{lapidoth}.

Considering the channel estimation results and interference $s$ caused by the primary users, we have the following achievable rate expressions as lower bounds to the instantaneous channel capacities in the above four scenarios:
\begin{align}\label{capacities}
\begin{split}
&\textit{Scenario 1:} \quad R_{1}=\frac{(T-N)B-1}{T}\log\left(1+\tSNR_{1}|w_{l}|^{2}\right)\\
&\textit{Scenario 2:} \quad R_{2}=\frac{(T-N)B-1}{T}\log\left(1+\tSNR_{2}|w_{l}|^{2}\right)\\
&\textit{Scenario 3:} \quad R_{3}=\frac{(T-N)B-1}{T}\log\left(1+\tSNR_{3}|w_{l}|^{2}\right)\\
&\textit{Scenario 4:} \quad R_{4}=\frac{(T-N)B-1}{T}\log\left(1+\tSNR_{4}|w_{l}|^{2}\right)
\end{split}
\end{align}
\vspace{-.3cm}
where
\begin{align}\label{snr}
\begin{split}
&\tSNR_{1}=\frac{P_{db}\sigma_{\widehat{h}_{l}}^{2}}{P_{db}\sigma_{\widetilde{h}_{1}}^{2}+\sigma_{n}^{2}+\sigma_{s}^{2}},\quad \tSNR_{2}=\frac{P_{di}\sigma_{\widehat{h}_{l}}^{2}}{P_{di}\sigma_{\widetilde{h}_{2}}^{2}+\sigma_{n}^{2}+\sigma_{s}^{2}}\\
&\tSNR_{3}=\frac{P_{db}\sigma_{\widehat{h}_{l}}^{2}}{P_{db}\sigma_{\widetilde{h}_{3}}^{2}+\sigma_{n}^{2}},\
\textrm{and}\
\tSNR_{4}=\frac{P_{di}\sigma_{\widehat{h}_{l}}^{2}}{P_{di}\sigma_{\widetilde{h}_{4}}^{2}+\sigma_{n}^{2}}.
\end{split}
\end{align}
These lower bounds are obtained by assuming that $\widetilde{h}_{l}x_{i}$ and $s_{i}$ are Gaussian distributed which is the worst-case noise. Above, we have defined $\widehat{h}_{l}=w_{l}\sigma_{\widehat{h}_{l}}$. Note that $w_{l}$ is a standard complex Gaussian random variable with zero mean and unit variance, i.e., $w_{l} \sim \mathcal{CN}(0,1)$, in mismatched and linear MMSE. Hence, $z_{l}=|w_{l}|^{2}$ has an exponential distribution with mean 1. $P_{db}$ and $P_{di}$ are the data symbols powers when the channel is busy and idle, respectively, and they can be written as
\begin{equation}\label{data power}
P_{db}=\frac{\overline{P}_{1}T-P_{tb}}{(T-N)B-1}\quad\textrm{and}\quad
P_{di}=\frac{\overline{P}_{2}T-P_{ti}}{(T-N)B-1}.
\end{equation}

While the receiver attempts to learn the channel through training, we assume that the transmitter is unaware of the channel conditions and transmits the
information at fixed rates $r_{1}$ and $r_{2}$,
depending on the channel being sensed as busy or idle, respectively. Therefore, the transmission rate is $r_1$ in scenarios 1 and 3, and $r_2$ in scenarios 2 and 4. If these rates are below the achievable rate expressions provided in (\ref{capacities}), i.e., if $r_{1}<R_{1},R_{3}$ or $r_{2}<R_{2},R_{4}$, the transmission is considered to be in the ON state and reliable communication is achieved at these rates. On the other hand, if $r_{1}\geq R_{1},R_{3}$ or $r_{2}\geq R_{2},R_{4}$, then we assume that outage occurs and reliable communication can not be achieved. In such a case, the channel is in the OFF state. To ensure the reception of correct data, a simple automatic repeat request (ARQ) mechanism needs to be incorporated in the communication protocol in the OFF state.

\begin{figure}
\begin{center}
\includegraphics[width =\figsize\textwidth]{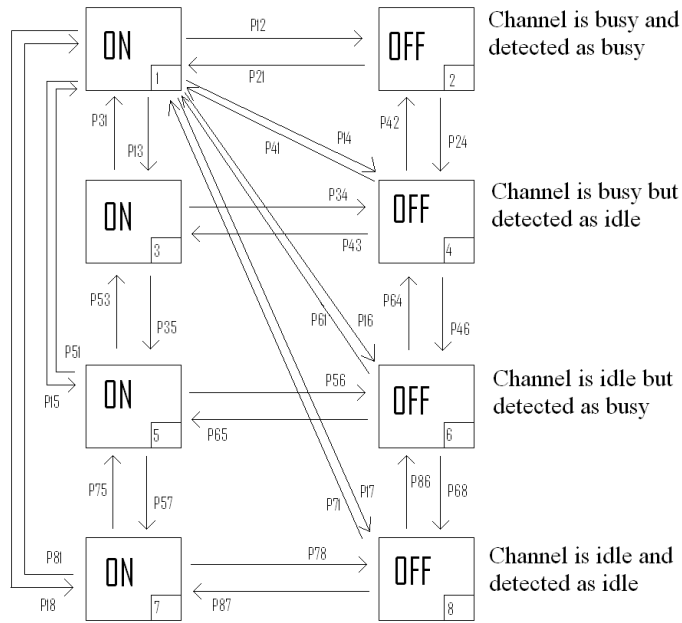}
\caption{State transition model for the cognitive radio channel. The numbered label for each state is given on the bottom-right corner of the box representing the state.} \label{fig:fig2}
\end{center}
\end{figure}

\begin{figure}
\begin{center}
\includegraphics[width =\figsize\textwidth]{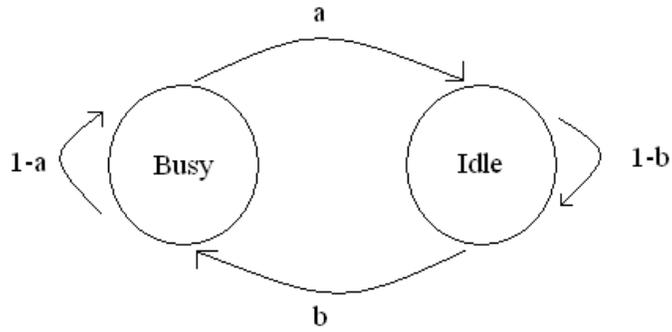}
\caption{Two-state Markov model for the primary user activity.}
\label{fig:fig1}
\end{center}
\end{figure}

From the above discussion, we see that we in each scenario have two states, namely ON and OFF, depending on whether or not the
fixed-transmission rate exceeds the instantaneous rate that the channel can support. Therefore, overall we have eight states.
Fig. \ref{fig:fig2} depicts the state transition model for the cognitive radio transmission considering all possible scenarios related to the channel sensing decisions and their correctness, and the reliability of the transmissions. The labels for the states are provided on the bottom-right corner of the box representing the state.

The transition probabilities in this state-transition model depend on the channel fading coefficients, the fixed transmission rates, and the primary user activity. Recall that we consider block-fading and assume that the fading coefficients stay constant throughout the frame and change independently from one frame to another. We also assume that primary user activity does not change within each frame. However, we employ a two-state Markov model to describe the transition of the primary user activity between the frames. This Markov model is depicted in Fig. \ref{fig:fig1}. Busy state indicates that the channel is occupied by the primary users, and idle state indicates that there is no primary user present in the channel. Probability of transitioning from busy state to idle state is denoted by $a$, and the probability of transitioning from idle state to busy state is denoted by $b$. Note that, by our assumption, state transitions happen every $T$ seconds, which is the frame duration.

Next, we determine the state transition probabilities. Let us first consider in detail the probability of staying in the topmost ON state in Fig. \ref{fig:fig2}. This probability, denoted by $p_{11}$, is given by
\begin{align}
p_{11}&= \Pr\left\{ \substack{\text{channel is busy and is detected busy,} \\ \text{and $r_1 < R_{1}(l)$ in the $l^{th}$ frame}} \,\, \Big | \,\, \substack{\text{channel is busy and is detected busy,} \\ \text{and $r_1 < R_{1}(l-1)$ in the $(l-1)^{th}$ frame}} \right\} \label{pro11_1}
\\
&= \Pr\left\{ \substack{\text{channel is busy } \\ \text{in the $l^{th}$ frame}} \,\, \Big | \,\, \substack{\text{channel is busy } \\ \text{in the $(l-1)^{th}$ frame}} \right\} \times \Pr\left\{ \substack{\text{channel is detected busy} \\ \text{in the $l^{th}$ frame}} \,\, \Big | \,\, \substack{\text{channel is busy} \\ \text{in the $l^{th}$ frame}} \right\} \nonumber
\\
& \hspace{.5cm}\times \Pr\left\{r_{1}<R_{1}(l) \mid r_{1}<R_{1}(l-1)\right\} \label{pro11_2}
\\
&=(1-a) \, P_{d} \, \Pr\left\{r_{1}<R_{1}(l)\mid r_{1}<R_{1-1}(l)\right\} \label{pro11_3}
\\
&=(1-a)\,P_{d}\,\Pr\left\{z_{l}>\alpha_{1}\mid
z_{l-1}>\alpha_{1}\right\} \label{pro11_4}
\\
&=(1-a)\,P_{d}\,\Pr\left\{z_{l}>\alpha_{1}\right\} \label{pro11_5}
\\
&=(1-a)\,P_{d}\,\Pr\left\{z>\alpha_{1}\right\} \label{pro11_6}
\end{align}
where
\begin{equation}\label{alfa1}
\alpha_{1}=\frac{2^{\frac{r_{1}T}{(T-N)B-1}}-1}{\tSNR_{1}},
\end{equation}
$P_{d}$ is the probability of detection in channel sensing, $r_1$ is the fixed transmission rate in scenario 1, and $R_1(l)$ denotes the achievable rate expression in scenario 1 in the $l^{th}$ frame. Above, (\ref{pro11_2}) is obtained by using the chain rule of probability\footnote{Consider the events $A,B,C,$ and $D$. Using the chain rule, the conditional probability $\Pr(A \bigcap B\bigcap C\bigcap \mid D)$ can be written as $\Pr(A \bigcap B\bigcap C\bigcap \mid D) = \Pr(A \mid D) \times \Pr(B \mid A \bigcap D) \times \Pr(C \mid A \bigcap B \bigcap D)$.} and noting the following facts. Channel being busy in the $l^{th}$ frame depends only on channel being busy in the $(l-1)^{th}$ frame and not on the other events in the condition. Moreover, since channel sensing is performed individually in each frame without any dependence on the channel sensing decision and primary user activity in the previous frame, channel being detected as busy in the $l^{th}$ frame depends only on the event that the channel is actually busy in the $l^{th}$ frame. Finally, the event $\{r_1 < R_{1}(l)\}$ is related to the channel fading coefficients and hence possibly depends on the event $\{r_1 < R_{1}(l-1)\}$ through the dependence of fading coefficients between frames. (\ref{pro11_3}) follows by realizing that the first probability in (\ref{pro11_2}) is equal to $(1-a)$, the probability of staying in the busy state in the Markov model given for primary user activity, and noticing that the second probability is equal to $P_d$, the  detection probability in channel sensing.
(\ref{pro11_4}) is obtained by noting that the event $\{r_{1}<R_{1}(l) = \frac{(T-N)B-1}{T}\log\left(1+\tSNR_{1}|w_{l}|^{2}\right)\}$ is equivalent to the event $\{z_l > \alpha_1\}$ where $z_l = |w_l|^2$ and $\alpha_1$ is defined in (\ref{alfa1}). (\ref{pro11_5}) follows from the fact that $z_l$ and $z_{l-1}$ are independent due to the block-fading assumption. Finally, (\ref{pro11_6}) is obtained by noting that fading coefficients and their estimates are identically distributed in each frame and hence the index $l$ in $z_l$ can be dropped.


Similarly, the probabilities for transitioning from any state to state 1 (topmost ON state) can be expressed as
\begin{align}\label{p11ler}
\begin{split}
p_{b1}=p_{11}&=p_{21}=p_{31}=p_{41}=(1-a)P_{d}\Pr\left\{z>\alpha_{1}\right\},\\
p_{i1}=p_{51}&=p_{61}=p_{71}=p_{81}=bP_{d}\Pr\left\{z>\alpha_{1}\right\}.
\end{split}
\end{align}
Note that we have common expressions for the transition probabilities in cases in which the originating state has a busy channel (i.e., states 1,2,3, and 4 ) and in cases in which the originating state has an idle channel (i.e., states 5,6,7, and 8).

In a similar manner, the remaining transition probabilities are given by the following:

For all $b \in \{1,2,3,4\}$ and  $i \in \{5,6,7,8\}$,
\begin{align}\label{prob2}
\begin{array}{ll}
p_{b2}=(1-a)P_{d}\Pr\left\{z\leq\alpha_{1}\right\}, \quad
&p_{i2}=bP_{d}\Pr\left\{z\leq\alpha_{1}\right\},\\
p_{b3}=(1-a)(1-P_{d})\Pr\left\{z>\alpha_{2}\right\},
&p_{i3}=b(1-P_{d})\Pr\left\{z>\alpha_{2}\right\},\\
p_{b4}=(1-a)(1-P_{d})\Pr\left\{z\leq\alpha_{2}\right\},
&p_{i4}=b(1-P_{d})\Pr\left\{z\leq\alpha_{2}\right\},\\
p_{b5}=aP_{f}\Pr\left\{z>\alpha_{3}\right\},
&p_{i5}=(1-b)P_{f}\Pr\left\{z>\alpha_{3}\right\},\\
p_{b6}=aP_{f}\Pr\left\{z\leq\alpha_{3}\right\},
&p_{i6}=(1-b)P_{f}\Pr\left\{z\leq\alpha_{3}\right\},\\
p_{b7}=a(1-P_{f})\Pr\left\{z>\alpha_{4}\right\},
&p_{i7}=(1-b)(1-P_{f})\Pr\left\{z>\alpha_{4}\right\},\\
p_{b8}=a(1-P_{f})\Pr\left\{z\leq\alpha_{4}\right\},
&p_{i8}=(1-b)(1-P_{f})\Pr\left\{z\leq\alpha_{4}\right\},\\
\end{array}
\end{align}
where $\alpha_{2}=\frac{2^{\frac{r_{2}T}{(T-N)B-1}}-1}{\tSNR_{2}}$,
$\alpha_{3}=\frac{2^{\frac{r_{1}T}{(T-N)B-1}}-1}{\tSNR_{3}}$, and
$\alpha_{4}=\frac{2^{\frac{r_{2}T}{(T-N)B-1}}-1}{\tSNR_{4}}$.
Note that since $b \in \{1,2,3,4\}$ is the index of the states with busy channels, we above have, for instance, $p_{b2} = p_{12} = p_{22} = p_{32} = p_{42}$.

Now, we can easily see that the $8\times8$ state transition matrix can be expressed as
\begin{equation}\label{R}
R=\left[
  \begin{array}{cccc}
    p_{1,1} & . & . & p_{1,8} \\
    . &   &   & . \\
    p_{4,1} & . & . & p_{4,8} \\
    p_{5,1} & . & . & p_{5,8} \\
    . &   &   & . \\
    p_{8,1} & . & . & p_{8,8} \\
  \end{array}
 \right]=
 \left[
  \begin{array}{cccc}
    p_{b1} & . & . & p_{b8} \\
    . &   &   & . \\
    p_{b1} & . & . & p_{b8} \\
    p_{i1} & . & . & p_{i8} \\
    . &   &   & . \\
    p_{i1} & . & . & p_{i8} \\
  \end{array}
 \right].
\end{equation}
Note that $R$ has a rank of 2.  Finally, we also note that $Tr_{1}$ and $Tr_{2}$ bits are
transmitted and received in the ON states 1 and 5, and 3 and 7,
respectively, while the transmitted number of bits is assumed to be
zero in the OFF states (i.e., in states 2, 4, 6, and 8).

\section{Effective Capacity}\label{Effective Capacity sec}
In this section, we identify the maximum throughput that the
cognitive radio channel with the aforementioned state-transition
model can sustain under statistical QoS constraints imposed in the
form of buffer or delay violation probabilities. Wu and Negi in
\cite{Wu} defined the effective capacity as the maximum constant
arrival rate that can be supported by a given channel service
process while also satisfying a statistical QoS requirement
specified by the QoS exponent $\theta$. If we define $Q$ as the
stationary queue length, then $\theta$ is defined as the decay rate
of the tail distribution of the queue length $Q$:
\begin{equation}\label{decayrate}
\lim_{q\rightarrow \infty}\frac{\log \Pr(Q\geq q)}{q}=-\theta.
\end{equation}
Hence, we have the following approximation for the buffer violation
probability for large $q_{max}$: $\Pr(Q\geq q_{max})\approx e^{-\theta
q_{max}}$. Therefore, larger $\theta$ corresponds to more strict QoS
constraints, while the smaller $\theta$ implies looser constraints.
In certain settings, constraints on the queue length can be linked
to limitations on the delay and hence delay-QoS constraints. It is
shown in \cite{Liu} that $\Pr\{D\geq d_{max}\}\leq c\sqrt{\Pr\{Q\geq
q_{max}\}}$ for constant arrival rates, where $D$ denotes the
steady-state delay experienced in the buffer. In the above
formulation, $c$ is a positive constant, $q_{max}=gd_{max}$ and $g$
is the source arrival rate. Therefore, \emph{effective capacity provides
the maximum arrival rate when the system is subject to statistical
queue length or delay constraints  in the forms of $\Pr(Q \ge
q_{\max}) \le e^{-\theta q_{max}}$ or $\Pr\{D \ge d_{\max}\} \le c \,
e^{-\theta g \, d_{max}/2}$, respectively}. Since the average arrival
rate is equal to the average departure rate when the queue is in
steady-state \cite{ChangZajic}, \emph{effective capacity can also be seen
as the maximum throughput in the presence of such constraints.}


The effective capacity for a given QoS exponent $\theta$ is given by
\begin{equation}\label{exponent}
-\lim_{t\rightarrow \infty}\frac{1}{\theta
t}\log_{e}\mathbb{E}\{e^{-\theta
S(t)}\} \triangleq -\frac{\Lambda(-\theta)}{\theta}
\end{equation}
where $\Lambda(\theta) = \lim_{t\rightarrow
\infty}\frac{1}{t}\log_{e}\mathbb{E}\{e^{\theta S(t)}\}$ is a
function that depends on the logarithm of the moment generating
function of $S(t)$, $S(t)=\sum_{k=1}^{t}r(k)$ is the
time-accumulated service process, and $\{r(k),k=1,2,\dots\}$ is
defined as the discrete-time, stationary and ergodic stochastic
service process. Note that the service rate is $r(k) = Tr_{1}$ if
the cognitive system is in state 1 or 5 at time $k$. Similarly, the
service rate is $r(k) = Tr_{2}$ in states 3 and 7. In all the OFF
states, fixed transmission rates exceed the instantaneous achievable rates, and outage occurs. Therefore,
the service rates in these states are effectively zero.

In the next result, we provide the effective capacity for the
cognitive radio channel and state transition model described in the
previous section.
\begin{theo} \label{theo:fixedrate}
For the cognitive radio channel with the state transition model given in Section \ref{sec:state}, the normalized effective capacity in bits/s/Hz is given by
\begin{align}\label{effective capacity}
\begin{split}
&R_{E}(\tSNR,\theta)=\max_{\substack{r_{1},r_{2}\geq0\\ \p_1 \le \p_{peak} \text{ and } \p_2 \le \p_{peak} \\P_{d}\overline{P}_{1}+(1-P_{d})\overline{P}_{2}\leq\overline{P}_{avg}}}-\frac{1}{\theta
TB} \log_{e}\\&\frac{1}{2}\bigg[(p_{b1}+p_{i5})e^{-\theta
Tr_{1}}+(p_{b3}+p_{i7})e^{-\theta
Tr_{2}}+p_{b2}+p_{b4}+p_{i6}+p_{i8}\bigg]\\&+\frac{1}{2}\bigg\{\bigg[(p_{b1}-p_{i5})e^{-\theta
Tr_{1}}+(p_{b3}-p_{i7})e^{-\theta
Tr_{2}}+p_{b2}+p_{b4}-p_{i6}-p_{i8}\bigg]^{2}\\&+4\left(p_{i1}e^{-\theta
Tr_{1}}+p_{i3}e^{-\theta
Tr_{2}}+p_{i2}+p_{i4}\right)\left(p_{b5}e^{-\theta
Tr_{1}}+p_{b7}e^{-\theta
Tr_{2}}+p_{b6}+p_{b8}\right)\bigg\}^{\frac{1}{2}}
\end{split}
\end{align}
where $T$ is the frame duration over which the fading stays
constant, $r_1$ and $r_2$ are fixed transmission rates, and $p_{bk}$
and $p_{ik}$ for $k = 1, \ldots, 8$, $b = 1,2,3,4$, and $i = 5,6,7,8$ are the transition probabilities
expressed in \eqref{p11ler} and \eqref{prob2}.
\end{theo}

\emph{Proof:} In \cite[Chap. 7, Example 7.2.7]{Performance}, it is
shown for Markov modulated processes that
\begin{gather} \label{eq:theta-envelope}
\frac{\Lambda(\theta)}{\theta} = \frac{1}{\theta} \log_e
sp(\phi(\theta)R)
\end{gather}
where $sp(\phi(\theta)R)$ is the spectral radius or the maximum of the absolute values of the eigenvalues of the matrix
$\phi(\theta)R$, $R$ is the transition matrix of the underlying Markov process, and $\phi(\theta) = \text{diag}(\phi_1(\theta), \ldots, \phi_M(\theta))$ is a diagonal matrix whose components are the moment generating functions of the processes in $M$ states ($M = 8$ in our case). The rates supported by the cognitive radio channel with the state transition model described in the previous section can be seen as a Markov modulated process and hence the setup considered in \cite{Performance} can be immediately applied to our setting. Note that the transmission rates are non-random and fixed in each state in the cognitive channel. More specifically, the possible rates are $Tr_{1}$, $Tr_{2}$, and 0 for which the moment generating functions are $e^{\theta Tr_1}$, $e^{\theta Tr_2}$, and 1, respectively.
Therefore, we have $\phi(\theta) = \text{diag}\{e^{\theta
Tr_{1}},1,e^{\theta Tr_{2}},1,e^{\theta Tr_{1}},1,e^{\theta
Tr_{2}},1\}$. Then, using \eqref{R}, we can write
\begin{align}
\phi(\theta)R=\left[
  \begin{array}{cccc}
    \phi_{1}(\theta)p_{b1} & . & . & \phi_{1}(\theta)p_{b8} \\
    . &   &   & . \\
    \phi_{3}(\theta)p_{b1} & . & . & \phi_{3}(\theta)p_{b8} \\
    \phi_{4}(\theta)p_{b1} & . & . & \phi_{4}(\theta)p_{b8} \\
    \phi_{5}(\theta)p_{i1} & . & . & \phi_{5}(\theta)p_{i8} \\
    . &   &   & . \\
    \phi_{8}(\theta)p_{i1} & . & . & \phi_{8}(\theta)p_{i8} \\
  \end{array}
 \right].
\end{align}
Since $\phi(\theta)R$ is a matrix with rank 2, we can readily find that \cite{onemli}
\begin{align}
sp(\phi(\theta)R) &=
\frac{1}{2}\bigg[\phi_{1}(\theta)p_{b1}
+...+\phi_{4}(\theta)p_{b4}+\phi_{5}(\theta)p_{i5}+...+
\phi_{8}(\theta)p_{i8}\bigg]\nonumber
\\&\hspace{.4cm}+\frac{1}{2}\Bigg\{\bigg[\phi_{1}(\theta)p_{b1}
+...+\phi_{4}(\theta)p_{b4}-\phi_{5}(\theta)p_{i5}-...-
\phi_{8}(\theta)p_{i8}\bigg]^{2}\nonumber
\\&\hspace{1.5cm}+4\Big(\phi_{1}(\theta)p_{i1}
+...+\phi_{4}(\theta)p_{i4}\Big)\Big(\phi_{5}(\theta)p_{b5}
+...+\phi_{8}(\theta)p_{b8}\Big) \Bigg\}^{\frac{1}{2}}\nonumber
\\
&=\frac{1}{2}\bigg[(p_{b1}+p_{i5})e^{\theta
Tr_{1}}+(p_{b3}+p_{i7})e^{\theta
Tr_{2}}+p_{b2}+p_{b4}+p_{i6}+p_{i8}\bigg]\nonumber
\\&\hspace{.4cm}+\frac{1}{2}\Bigg\{\bigg[(p_{b1}-p_{i5})e^{\theta
Tr_{1}}+(p_{b3}-p_{i7})e^{\theta
Tr_{2}}+p_{b2}+p_{b4}-p_{i6}-p_{i8}\bigg]^{2}\nonumber
\\&\hspace{1.5cm}+4\bigg(p_{i1}e^{\theta
Tr_{1}}+p_{i3}e^{\theta
Tr_{2}}+p_{i2}+p_{i4}\bigg)\bigg(p_{b5}e^{\theta
Tr_{1}}+p_{b7}e^{\theta
Tr_{2}}+p_{b6}+p_{b8}\bigg)\Bigg\}^{\frac{1}{2}}. \label{eq:sp}
\end{align}
Then, combining (\ref{eq:sp}) with (\ref{eq:theta-envelope}) and
(\ref{exponent}), we obtain the expression inside the maximization
on the right-hand side of (\ref{effective capacity}). Note that this
expression is the effective capacity for given values of fixed
transmission rates $r_{1}$ and $r_{2}$ and of average power levels
$\overline{P}_{1}$ and $\overline{P}_{2}$, and can be maximized by
choosing the optimal values of $r_1$ and $r_2$ over the optimized
power allocation policy. This maximization leads to the effective
capacity formula given in (\ref{effective capacity}). Note also that we have normalized the effective capacity expression in (\ref{effective capacity}) by $TB$ to have it in the units of bits/s/Hz. \hfill
$\square$

\vspace{-.3cm}
\section{Numerical Results}\label{Numerical Results}

In this section, we present the numerical results. In our
simulations, we assume that the fading coefficients are zero-mean Gaussian random variables with unit variance, $\sigma_{h}^{2}=1$. Note also that in all of the simulations we assume $T=0.1$ seconds, $N=0.01$ seconds, $B=1000$ Hz, $\sigma_{n}^{2}=1$, $\sigma_{s}^{2}=1$, $\theta=0.1$, $a=0.9$, $b=0.1$, and $\eta=0.1$. Unless stated otherwise, we assume in the numerical results that mismatch MMSE with estimates given in (\ref{eq:estimate1}) -- (\ref{eq:estimate4}) is employed in the training phase. Moreover, we set $\p_{peak} = 10$ dB.

\begin{figure}
\begin{center}
\includegraphics[width =\figsize\textwidth]{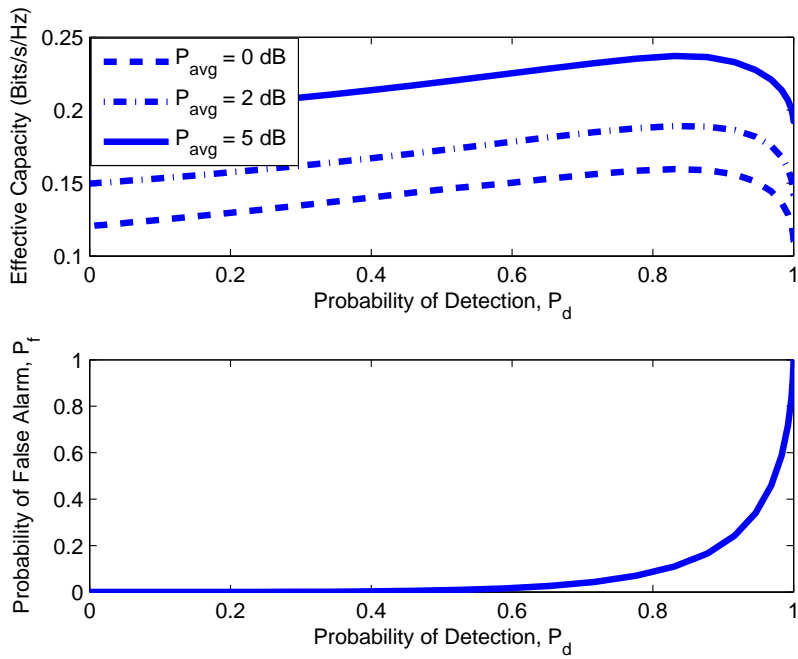}
\caption{Upper Figure: Effective capacity vs. detection probability $P_d$ for different values of $\bar{P}_{avg}$. Lower Figure: False alarm probability $P_f$ vs. $P_d$.} \label{fig:fig3}
\end{center}
\end{figure}

\begin{figure}
\begin{center}
\includegraphics[width =\figsize\textwidth]{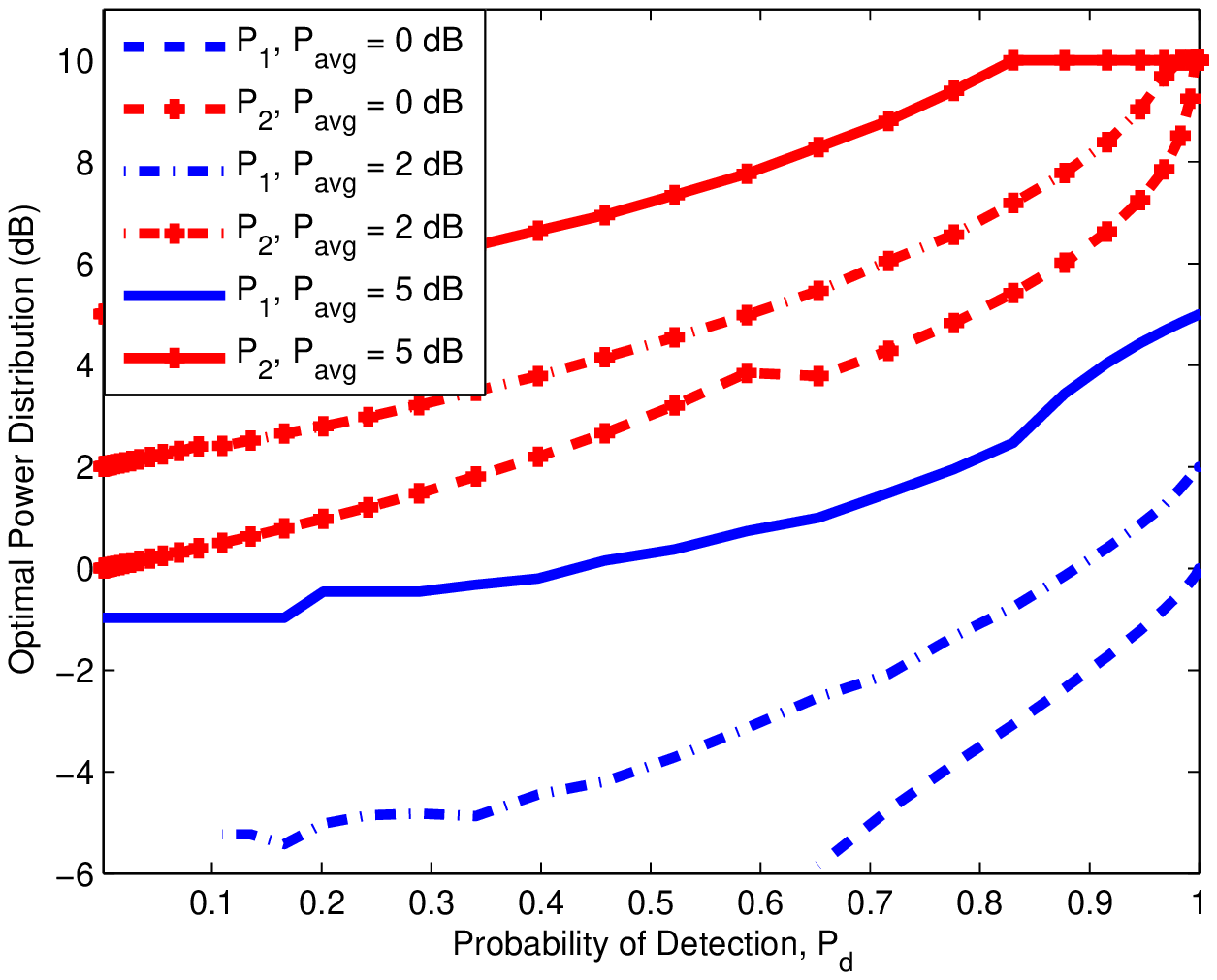}
\caption{Optimal values of $\bar{P}_1$ and $\bar{P}_2$ vs. detection probability $P_d$ for different values of $\bar{P}_{avg}$.} \label{fig:fig4}
\end{center}
\end{figure}

In Figure \ref{fig:fig3}, we display the optimal effective capacity as a function of the probability of detection, $P_{d}$, for different values of $\overline{P}_{avg}$. As expected, with increasing $\overline{P}_{avg}$, the effective capacity value increases. Note also that probability of false alarm $P_{f}$ is displayed in the second half of Fig. \ref{fig:fig3}. It is clear that the maximum effective capacity values are obtained when $P_{d}$ is close to 0.9. As $P_d$ further increases and approaches 1, we notice in the lower plot in Fig. \ref{fig:fig3} that false-alarm probability increases to 1 as well. Hence, the secondary users start to regard the channel busy all the time and performance degradations are experienced  because of not being able to take advantage of idle channel states. In Fig. \ref{fig:fig4}, the optimal values of $\overline{P}_{1}$ and $\overline{P}_{2}$ for different values of $\overline{P}_{avg}$ are displayed again as a function of $P_d$. Recall that $\overline{P}_{1}$ and $\overline{P}_{2}$ are the transmission power levels when the channel is sensed as busy and idle, respectively. First, we note that generally the power levels increase with increasing detection probability values. Also, we see in the figure that $\overline{P}_{2}$ is generally larger than $\overline{P}_{1}$. Hence, more power is allocated to cases in which the channel is detected as idle. As $P_d$ increases, we note from (\ref{general power constraint}) that the constraint on $\overline{P}_{2}$ relaxes since $\overline{P}_{2}$ is multiplied by a smaller weight ($1-P_d$). Consequently, $\overline{P}_{2}$ increases. Indeed, as $P_d \to 1$, the only constraint on $\p_2$ is $\p_2 \le \p_{peak} = 10$ dB.  Hence, the optimal value is $\p_2 = \p_{peak}$, and we actually observe in the figure that all $\p_2$ curves converge to 10 dB as $P_d$ approaches 1. On the other hand, as $P_d \to 1$, (\ref{general power constraint}) becomes $\p_1 \le \p_{avg}$. Since $\p_{avg} < \p_{peak} = 10$ dB, the only active constraint on $\p_1$ is $\p_1 \le \p_{avg}$ and it is noted in the figure that $\p_1$ approaches the optimal value $\p_{avg}$ as $P_d$ increases to 1. On the other hand, we interestingly observe that for relatively low values of $\p_{avg}$ (e.g., $\p_{avg} = 0,2$ dB), we have $\p_1 = 0$ if $P_d$ is below a certain threshold. Hence, no transmission is performed when the channel is sensed as busy. As $P_d$ further decreases and approaches 0, the secondary users always miss the primary user activities, and (\ref{general power constraint}) becomes $\p_2 \le \p_{avg}$, which is, similarly as discussed above, is the only active constraint for $\p_2$. Indeed, $\p_2$ curves approach the corresponding $\p_{avg}$ values as $P_d \to 0$.

\begin{figure}
\begin{center}
\includegraphics[width =\figsize\textwidth]{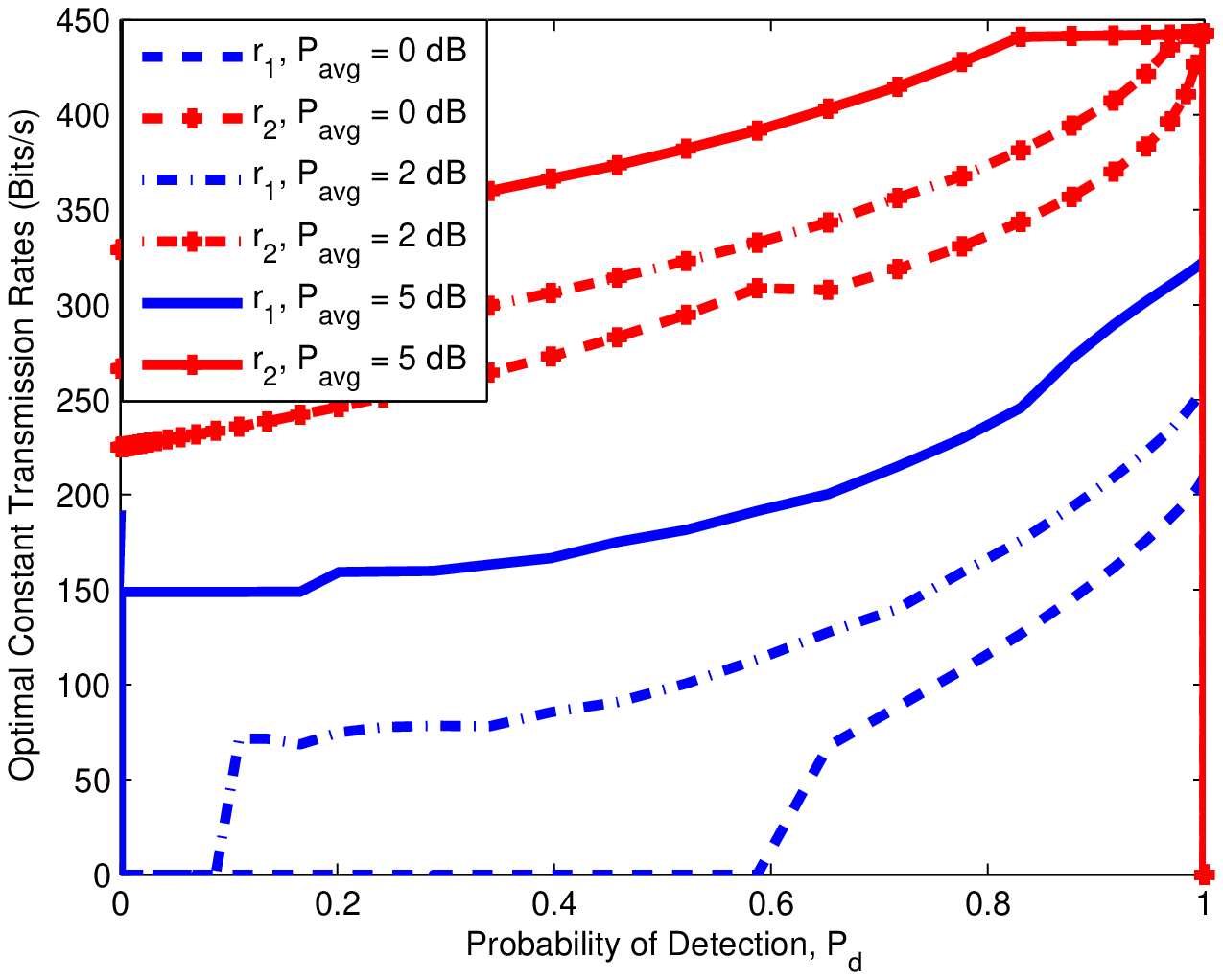}
\caption{Optimal values of $r_1$ and $r_2$ vs. detection probability $P_d$ for different values of $\bar{P}_{avg}$.} \label{fig:fig5}
\end{center}
\end{figure}

In Fig. \ref{fig:fig5}, we show the optimal fixed transmission rates $r_1$ and $r_2$ as a function of $P_{d}$ for different values of $\overline{P}_{avg}$. Note that the optimal transmission rates are obtained at optimal power levels. We observe that $r_{2}$, the transmission rate when the channel is detected as idle, is larger than $r_{1}$. In general, we note similar trends as in Fig. \ref{fig:fig4}.

\begin{figure}
\begin{center}
\includegraphics[width =\figsize\textwidth]{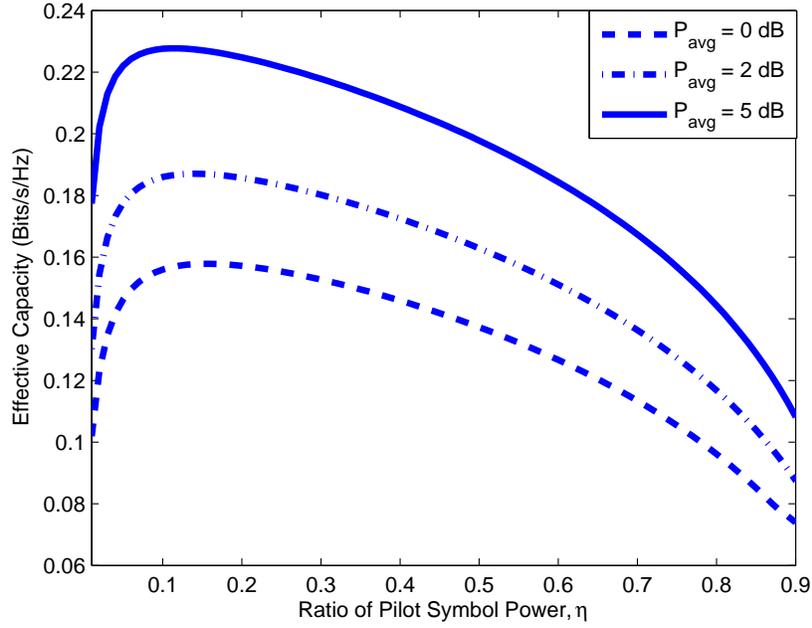}
\caption{Effective capacity vs. $\eta$, the fraction of total power allocated to the pilot symbol, for different values of $\bar{P}_{avg}$.} \label{fig:fig6}
\end{center}
\end{figure}

\begin{figure}
\begin{center}
\includegraphics[width =\figsize\textwidth]{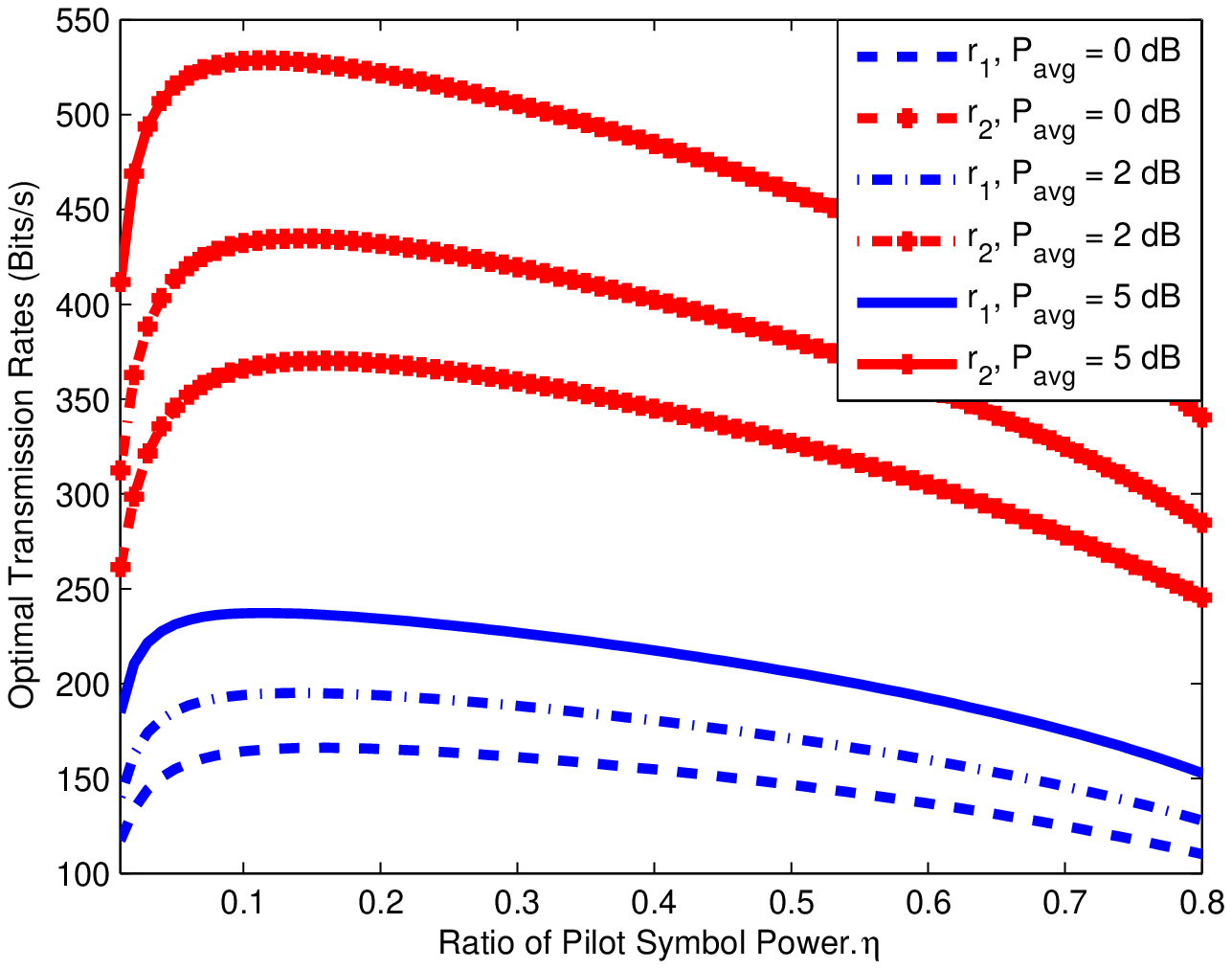}
\caption{Optimal values of $r_1$ and $r_2$ vs. $\eta$, the fraction of total power allocated to the pilot symbol, for different values of $\bar{P}_{avg}$.} \label{fig:fig7}
\end{center}
\end{figure}

\begin{figure}
\begin{center}
\includegraphics[width =\figsize\textwidth]{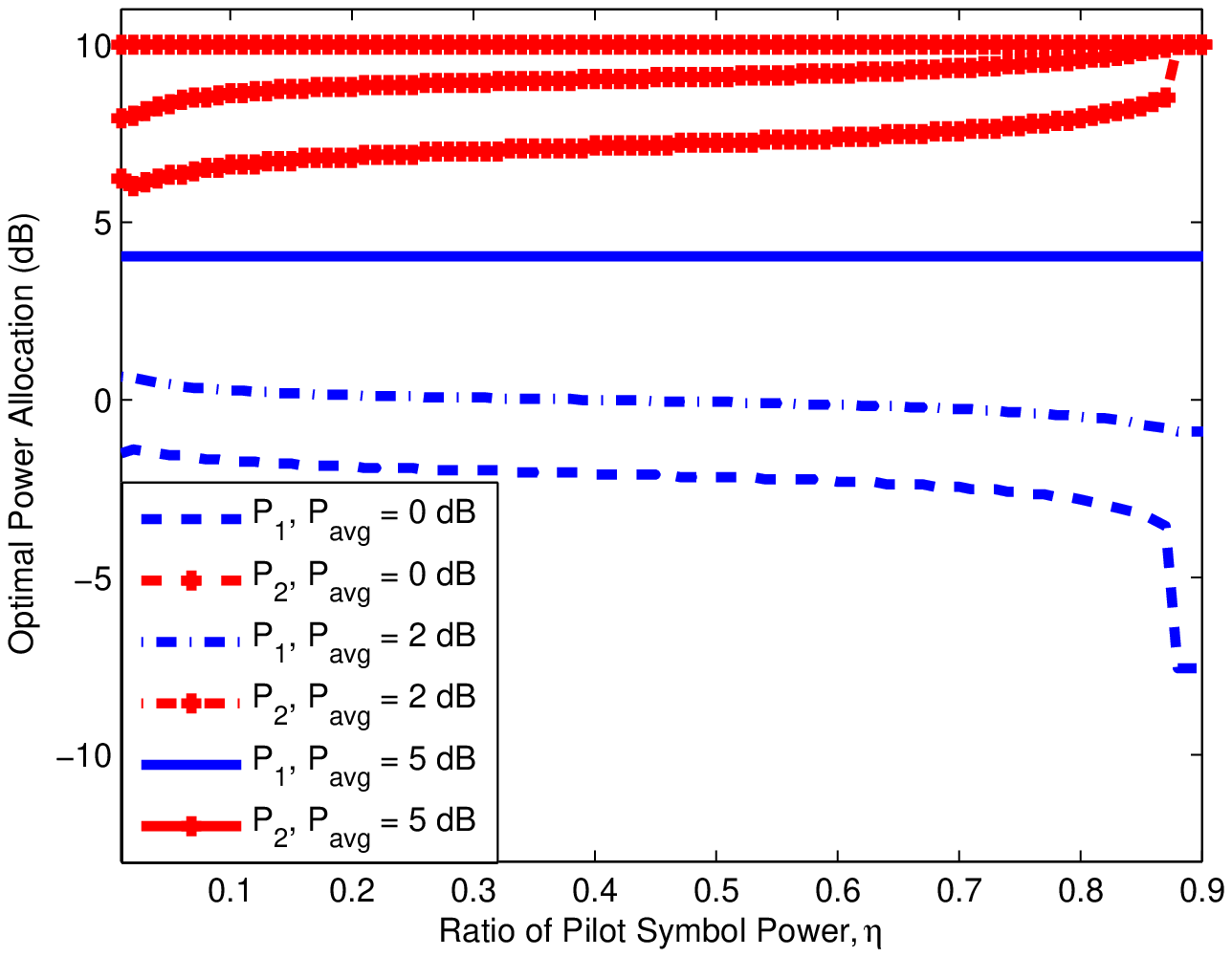}
\caption{Optimal values of $\bar{P}_1$ and $\bar{P}_2$  vs. $\eta$, the fraction of total power allocated to the pilot symbol, for different values of $\bar{P}_{avg}$.} \label{fig:fig8}
\end{center}
\end{figure}

\begin{figure}
\begin{center}
\includegraphics[width =\figsize\textwidth]{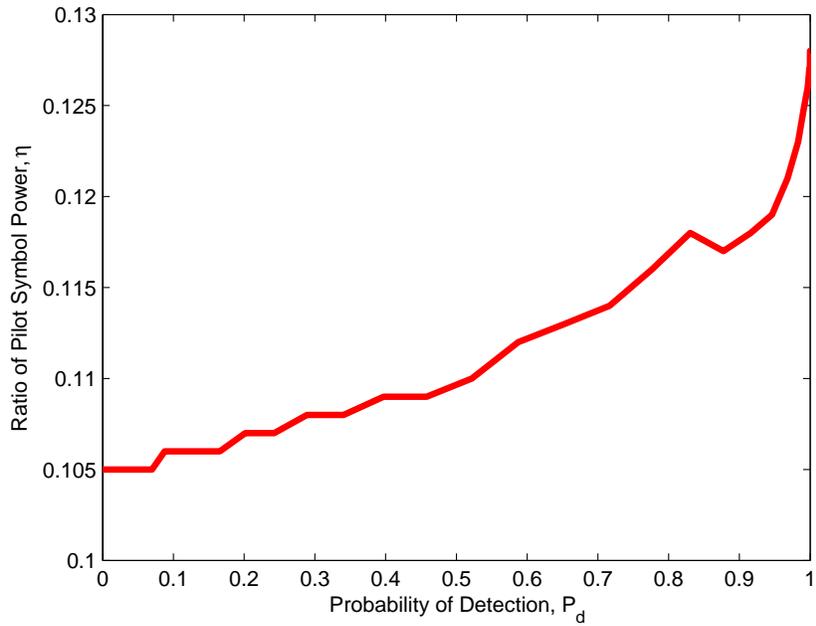}
\caption{$\eta$, the fraction of total power allocated to the pilot symbol, vs detection probability $P_d$. $\p_{avg} = 5$ dB.} \label{fig:eklenecek}
\end{center}
\end{figure}

In Fig. \ref{fig:fig6}, we plot the effective rate as a function of $\eta$, the fraction  of the total power allocated to the pilot symbol. We again consider three different average power levels. When the average power levels are 0, 2, and 5 dB, the optimal fractions are $\eta=0.16, 0.14$, and $0.11$, respectively. In this figure, we have $P_{d}=0.92$ and $P_{f}=0.24$. Hence, these are the optimal training power levels in the presence of channel sensing errors. In Fig. \ref{fig:fig7}, the optimal transmission rates are plotted as a function of $\eta$. It is observed that at all average power levels, the optimal transmission rate when the channel is sensed to be idle, i.e., $r_{2}$, is higher than the optimal transmission rate when the channel is detected as busy, i.e., $r_{1}$. In Fig. \ref{fig:fig8}, $P_{1}$ and $P_{2}$ are plotted as a function of $\eta$. It is observed that $P_{2}$ is higher than $P_{1}$ at all average power levels and for all values of $\eta$. Note that the optimal power distributions are obtained for constant $P_{d}$ and $P_{f}$. Finally, we plot in Fig. \ref{fig:eklenecek} $\eta$ vs. $P_d$, and observe that the fraction of power allocated to training increases with increasing $P_d$.

\begin{figure}
\begin{center}
\includegraphics[width =\figsize\textwidth]{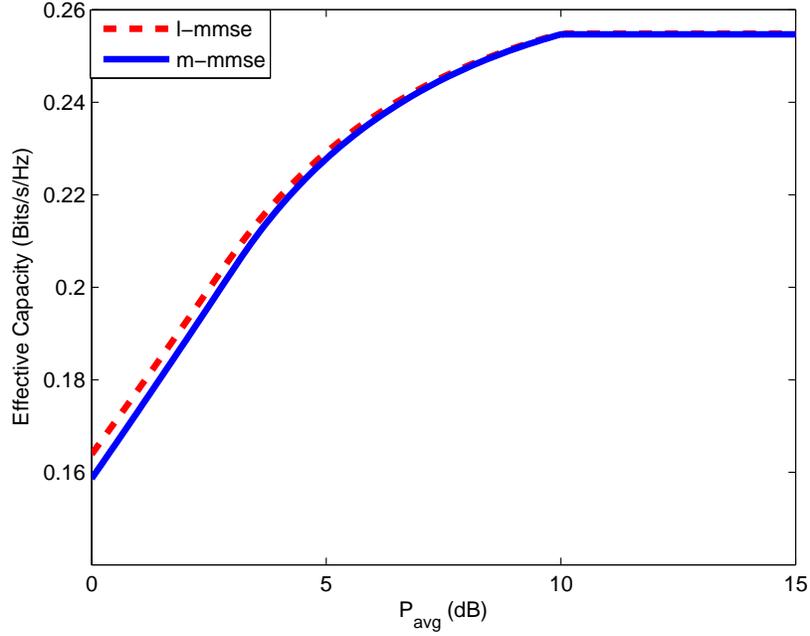}
\caption{Effective Capacity vs. $\bar{P}_{avg}$ when $m-mmse$ and $l-mmse$ estimation techniques are employed.} \label{fig:FigureComparison1}
\end{center}
\end{figure}

\begin{figure}
\begin{center}
\includegraphics[width =\figsize\textwidth]{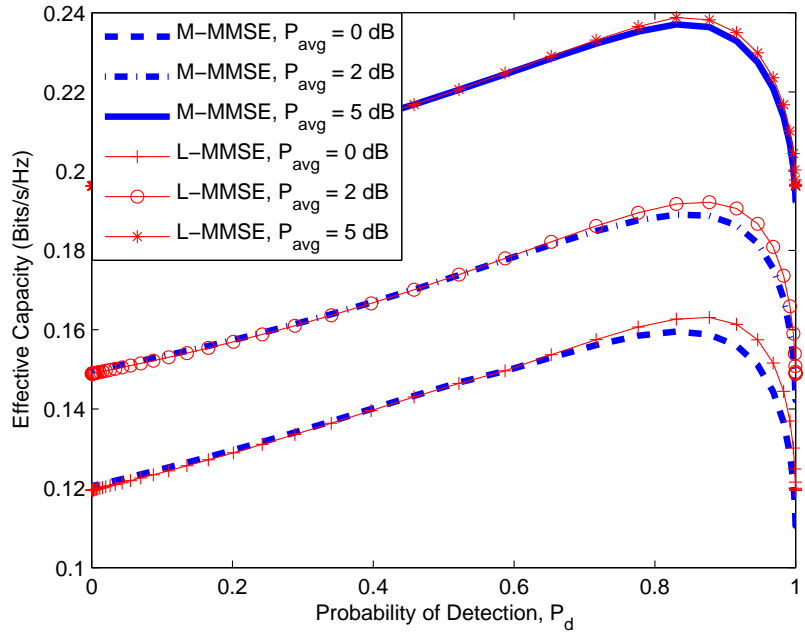}
\caption{Effective capacity vs. detection probability $P_d$ for different values of $\bar{P}_{avg}$ when $m-mmse$ and $l-mmse$ estimation techniques are employed.} \label{fig:FigureComparison2}
\end{center}
\end{figure}

Heretofore in the numerical results, we have assumed that mismatched MMSE is employed at the receiver.  In Figs. \ref{fig:FigureComparison1} and \ref{fig:FigureComparison2}, we compare the effective capacity values obtained using mismatched MMSE and linear MMSE techniques. In Fig. \ref{fig:FigureComparison1}, we plot the effective capacity vs. $\overline{P}_{avg}$ curve. We notice that linear MMSE provides a slightly better performance for low $\overline{P}_{avg}$. The performance gap vanishes as $\overline{P}_{avg}$ increases. In Fig. \ref{fig:FigureComparison2}, a similar conclusion is also reached. In this figure, we also observe that linear MMSE provides gains especially when the detection probability $P_d$ is high. Note that this is another interesting observation indicating the strong interactions between channel sensing and channel estimation.

\section{Conclusion}\label{Conclusion}
In this paper, we have analyzed the effective capacity of cognitive radio channels in the presence of QoS constraints, channel uncertainty, and transmission power limitations. We have considered a system model in which the cognitive SUs initially perform channel sensing and estimation, and subsequently transmit data. Channel sensing is done through energy detection and is formulated as a hypothesis testing problem. We have considered different estimation techniques, namely, mismatched MMSE, linear MMSE, and MMSE, in the training phase. In this setting, we have identified the interactions between channel sensing and estimation. In particular, we have noted that sensing errors lead to degradations in the estimation results. We have also shown that imperfections in sensing complicate MMSE estimation, and suboptimal techniques such as mismatched and linear MMSE enable tractable analysis.

In the data transmission phase, we have assumed that the transmitter, not being aware of the channel conditions, send the data at fixed power and rate. We have further assumed that these transmission parameters depend on whether the channel is sensed as busy or idle. For this cognitive operation, we have constructed a state-transition model by considering the reliability of the transmissions, channel sensing decisions and their correctness, and the evolution of primary user activity which is modeled as a two-state Markov process. We have formulated the transition probabilities in this model. Then, for the constructed state-transition model, we have obtained an expression for the effective capacity and identified the maximum throughput in the presence of buffer constraints. We have performed a numerical analysis and shown the impact of several parameters such as detection and false probabilities, average power constraints, training power value, on the performance. We have determined the optimal transmission power and rate levels. We have also compared the performances of linear and mismatched MMSE estimation methods.

\appendix

\subsection{Proof of Theorem \ref{theo:MMSEestimate}} \label{app:proofMMSEestimate}

In the cognitive scenario we are considering, the signal received by the receiver in the training phase is
\begin{gather}
y = \left\{
\begin{array}{ll}
\sqrt{P_t} h + n + s & \text{if the channel is busy}
\\
\sqrt{P_t} h + n & \text{if the channel is idle}
\end{array}\right.. \label{eq:trainingoutput}
\end{gather}
Note that we assume that $n$ and $s$ are independent complex Gaussian random variables with zero-mean and variances $\sigma_n^2$ and $\sigma_s^2$, respectively. Therefore, the variance of the noise component\footnote{Noise component is $n + s$ when the channel is busy, and $n$ when the channel is idle.} is either $\sigma_n^2 + \sigma_s^2$ or $\sigma_n^2$, depending on whether the channel is busy or idle. Since the receiver does not perfectly know the state of the primary user activity and only has a guess through channel sensing, the noise variance, $\sigma^2$, is random taking two values: $\sigma_n^2 + \sigma_s^2$ and $\sigma_n^2$. Now, the MMSE estimate in the presence of uncertainty in the noise statistics is obtained as follows:
\vspace{-.4cm}
\begin{align}
\widehat{h}_{mmse} &= \E\{h | y\}
\\
&= P(\sigma^2 = \sigma_n^2 \mid y) \E\{ h \mid  y, \sigma^2 = \sigma_n^2\} + P(\sigma^2 = \sigma_n^2 + \sigma_s^2 \mid y) \E\{ h \mid  y, \sigma^2 = \sigma_n^2 + \sigma_s^2\} \label{eq:mmseproof1}
\\
&= P\{\sigma^2 = \sigma_n^2 \mid y\} \, \frac{\sqrt{P_{t}}\sigma_{h}^{2}}{P_{t}\sigma_{h}^{2}+\sigma_{n}^{2}}y_{} + P\{\sigma^2 = \sigma_n^2 + \sigma_s^2 \mid y\} \, \frac{\sqrt{P_{t}}\sigma_{h}^{2}}{P_{t}\sigma_{h}^{2}+\sigma_{n}^{2}+\sigma_{s}^{2}}y_{}. \label{eq:mmseproof2}
\end{align}
Above, (\ref{eq:mmseproof1}) is obtained by using the following property of conditional expectation: $
\E\{X \mid Y \} = \E\{\, \E\{X \mid Y, Z\} \mid Y \}
$
where the outer expectation on the right-hand side is with respect to the conditional distribution of $Z$ given $Y$. In our setting, $Z$ is the noise variance. Hence, the above formulation indicates that we can find the MMSE estimate by evaluating the average of the MMSE estimates with fixed noise variances with respect to the conditional distribution of the noise variance given the observation. This is indeed what is done in (\ref{eq:mmseproof1}). (\ref{eq:mmseproof2}) is obtained by noting that once the noise variance is fixed, the MMSE estimates in a Gaussian setting are given by $\E\{ h \mid  y, \sigma^2 = \sigma_n^2\} = \frac{\sqrt{P_{t}}\sigma_{h}^{2}}{P_{t}\sigma_{h}^{2}+\sigma_{n}^{2}}y_{} $ and $\E\{ h \mid  y, \sigma^2 = \sigma_n^2 + \sigma_s^2\} = \frac{\sqrt{P_{t}}\sigma_{h}^{2}}{P_{t}\sigma_{h}^{2}+\sigma_{n}^{2}+\sigma_{s}^{2}}y_{} $.

Next, we provide the expressions for the conditional probabilities using Bayes' rule:
\begin{align*}
P\{\sigma^2 = \sigma_n^2 \mid y\} &= \frac{P\{\sigma^2 = \sigma_n^2\}f(y|\sigma^2 = \sigma_n^2)}{f(y)},
\text{ and }
P\{\sigma^2 = \sigma_n^2 + \sigma_s^2 \mid y\} = \frac{P\{\sigma^2 = \sigma_n^2 + \sigma_s^2\}f(y|\sigma^2 = \sigma_n^2 + \sigma_s^2)}{f(y)}.
\end{align*}
Given the value of the noise variance $\sigma^2$, $y$ is conditionally Gaussian distributed with zero mean and variance $\sigma^2$, as can be immediately seen from the relations in (\ref{eq:trainingoutput}). These conditional Gaussian distributions are provided in (\ref{eq:fy1}) and (\ref{eq:fy2}) in Section \ref{subsec:training}. $f(y)$ is the average of the conditional distributions and hence is given by  (\ref{eq:fyavg}). The prior probability of the noise variance depends on the channel sensing result. For instance, let us assume that the channel is detected as busy. Then,
\begin{align}
P\{\sigma^2 = \sigma_n^2\} &= P\{\substack{\text{channel} \\ \text{is idle}} \mid \substack{\text{channel} \\ \text{is detected busy}}\}
\\
&= \frac{P\{\substack{\text{channel}\\ \text{is idle}}\} P\{\substack{\text{channel is} \\ \text{detected busy}} \mid \substack{\text{channel} \\ \text{is idle}}\}}{P\{\substack{\text{channel is} \\ \text{detected busy}}\}}
\\
&= \frac{P\{\substack{\text{channel}\\ \text{is idle}}\} P\{\substack{\text{channel is} \\ \text{detected busy}} \mid \substack{\text{channel} \\ \text{is idle}}\}}{P\{\substack{\text{channel} \\ \text{is idle}}\} P\{\substack{\text{channel is} \\ \text{detected busy}} \mid \substack{\text{channel} \\
\text{is idle}}\} + P\{\substack{\text{channel} \\ \text{is busy}}\} P\{\substack{\text{channel is} \\ \text{detected busy}} \mid \substack{\text{channel} \\
\text{is busy}}\}}
\\
&= \frac{\frac{a}{a+b} P_f}{ \frac{a}{a+b} P_f + \frac{b}{a+b} P_d}
\\
&= \frac{a P_f}{a P_f + b P_d} \label{eq:Psigma=sigma_n}
\end{align}
Note that having $\sigma^2 = \sigma_n^2$ means that there are no primary users in the channel and hence channel is idle. By our assumption, channel is detected as busy. Therefore, $P\{\sigma^2 = \sigma_n^2\}$ is equal to the conditional probability $P\{\substack{\text{channel} \\ \text{is idle}} \mid \substack{\text{channel} \\ \text{is detected busy}}\}$. Then, the expression in (\ref{eq:Psigma=sigma_n}) is obtained by noting that $P\{\substack{\text{channel}\\ \text{is idle}}\} = \frac{a}{a+b}$ and $P\{\substack{\text{channel}\\ \text{is busy}}\} = \frac{b}{a+b}$, which can be derived easily from the two-state Markov chain used for primary user activity, and by realizing that $P\{\substack{\text{channel is} \\ \text{detected busy}} \mid \substack{\text{channel} \\ \text{is idle}}\}$ is the false alarm probability $P_f$ and $P\{\substack{\text{channel is} \\ \text{detected busy}} \mid \substack{\text{channel} \\
\text{is busy}}\}$ is the detection probability $P_d$.  The expressions in (\ref{eq:prior1}) and (\ref{eq:prior2}) for the other cases are obtained using a similar approach.
\end{spacing}
\vspace{-.3cm}
\begin{spacing}{1.1}

\end{spacing}

\newpage

\end{document}